\documentclass{moriond}

\def\Journal#1#2#3#4{{#1} {\bf #2}, #3 (#4)}

\def\PLB{{\em Phys. Lett.}  B}

\def\PRD{{\em Phys. Rev.} D}

\def\mco{\multicolumn}

\def\ra{\rightarrow}

\def\ko{K^0}

\def\be{\begin{equation}}
\def\ee{\end{equation}}
\def\bea{\begin{eqnarray}}
\def\eea{\end{eqnarray}}

\newcommand{\Photo}{}

\begin{document}
\vspace*{4cm}
\title{How in situ atmospheric transmission can affect cosmological 
constraints from type Ia supernovae ?}

\author{J\'er\'emy Neveu$^{1,2}$, Vincent Br\'emaud$^{1,2}$, S\'ebastien Bongard$^{1}$, Yannick Copin$^{3}$, Sylvie Dagoret-Campagne$^{2}$, Marc Moniez$^{2}$}

\address{$^{1}$Sorbonne Universit\'e, CNRS, Universit\'e de Paris, LPNHE, 75252 Paris Cedex 05, France; $^{2}$Universit\'e Paris-Saclay, CNRS, IJCLab, 91405, Orsay, France; $^{3}$Univ Lyon, Univ Claude Bernard Lyon 1, CNRS/IN2P3, IP2I Lyon, UMR 5822, F-69622, Villeurbanne, France.}

	
\maketitle\abstracts{
The measurement of type Ia supernova colours in photometric surveys is the key to access to cosmological distances. But for future large surveys like the Large Survey of Space and Time undertaken by the Vera Rubin Observatory in Chile, the large statistical power of the promised catalogues will make the photometric calibration uncertainties dominant in the error budget and will limit our ability to use it for precision cosmology. The knowledge of the on-site atmospheric transmission on average for the full survey, or for season or each exposure can help reaching the sub-percent precision for magnitudes. We will show that measuring the local atmospheric transmission allows to correct the raw magnitudes to reduce the photometric systematic uncertainties. Then we will present how this strategy is implemented at the Rubin Observatory via the Auxiliary Telescope and its slitless spectrograph. 
}


Cosmology measures and interprets the evolution of the whole universe. To probe its dynamic and understand the nature of dark energy, observers needs to compute distances at different epochs, from the light they received in telescopes. The evolution of cosmological distances with time tells how dark energy, dark matter and matter interacts and how they can be modelled. 

Optical surveys use magnitude and colour comparisons to build a relative distance scale. For instance, type Ia supernovae (SNe~Ia) revealed the presence of a dark energy component because they appeared fainter in the early universe than it was thought. More precisely, because SNIa colours redshift with universe expansion,  high redshift supernovae were fainter in red bands than what can be inferred from low redshift supernovae observed in blue bands. This case underlines that colours need to be accurately calibrated in an optical survey to display the universe dynamics. Every chromatic effect that alter the astral light distort our dynamic perception of the universe expansion, like the galactic dust, the instrumental response or the local atmospheric conditions.


Let's take the example of the future Large Survey of Space and Time undertaken by the Vera Rubin Observatory in Chile (Cerro Pachon). This is a 10 year optical survey using 6 broad bandpass filters $ugrizy$, that will experience many different atmospheric conditions. The measured flux $\Phi_b^{\mathrm{mes}}$ in a band $b$ and the corresponding magnitude $b^{\mathrm{mes}}$ are:
\begin{equation}
\Phi_b^{\mathrm{mes}} = \int_0^{\infty} S_*(\lambda) R_b(\lambda) T_{\mathrm{atm}}(\lambda \vert Z, t_{\mathrm{exp}}, \theta_a) \lambda \mathrm{d}\lambda / hc, \quad b^{\mathrm{mes}}  = -2.5            \log_{10} \left[\Phi_b^{\mathrm{mes}}\right] + ZP
\end{equation}
with $ZP$ is the reference zero point for the exposure, $S_*(\lambda)$ the source observer-frame SED, $R_b(\lambda)$ the LSST full instrumental response in band $b$ and $T_{\mathrm{atm}}(\lambda \vert Z, t_{\mathrm{exp}}, \theta_a)$ the local atmospheric transmission at airmass $Z$, date $t_{\mathrm{exp}}$ depending on some parameters $\theta_a$ computed with Libradtran~\cite{libradtran2005}. In this paper we only consider the Precipitable Water Vapor (PWV, in mm). 

\begin{figure}
\centering
\includegraphics[width=0.46\columnwidth]{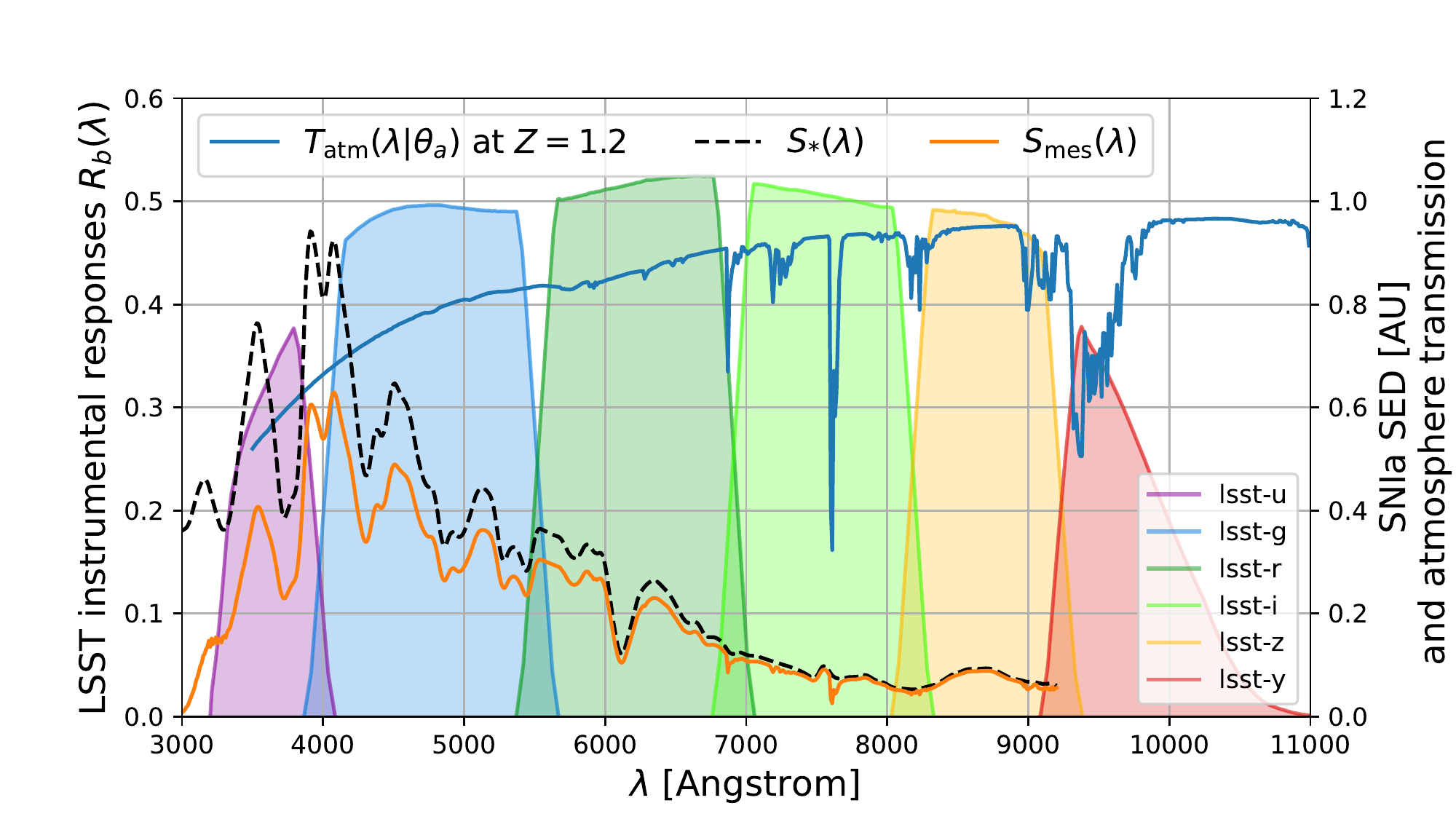}\hfill
\includegraphics[width=0.52\columnwidth]{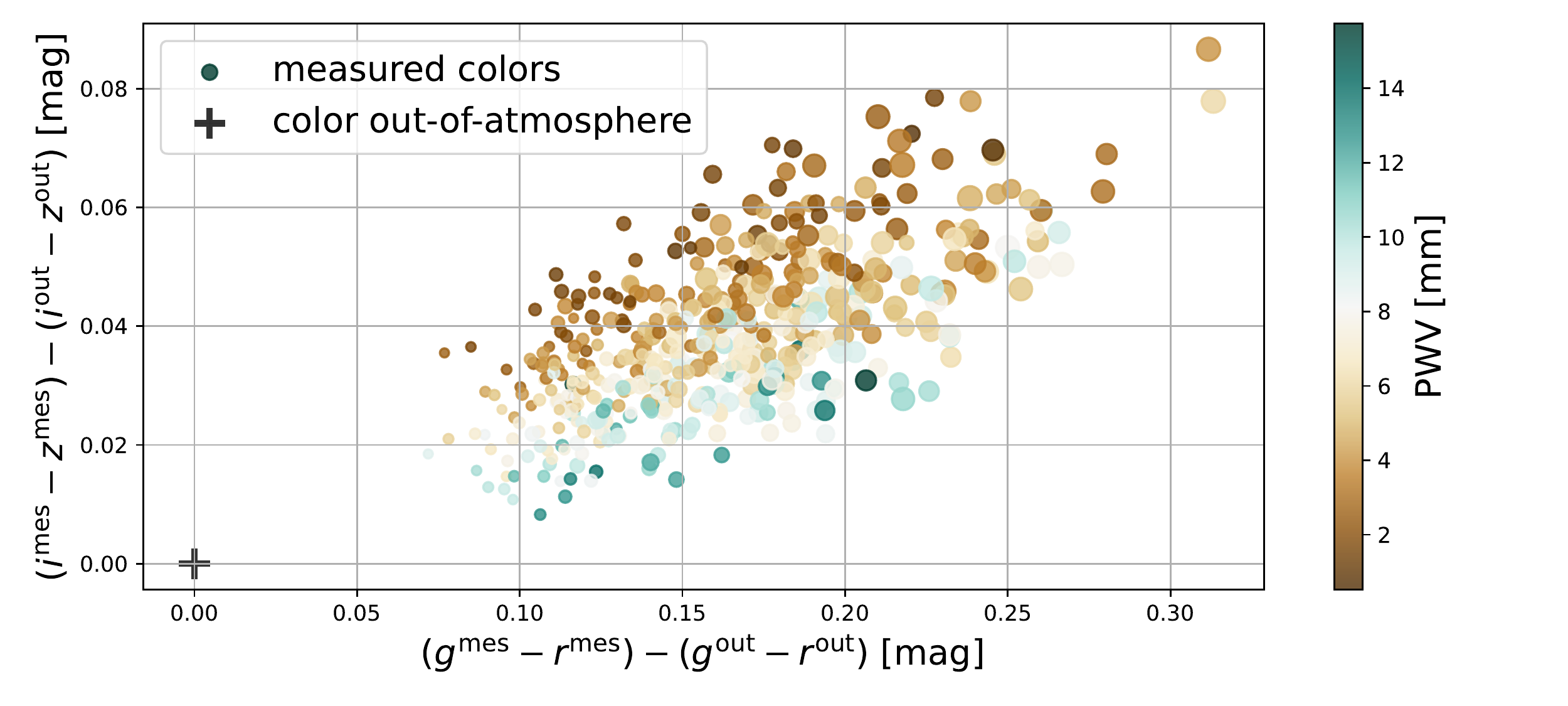}\\
\caption{\textit{Left:} LSST instrumental response per filter, typical SNIa SED $S_*(\lambda)$ at maximum, mean $T_{\mathrm{atm}}(\lambda)$ for LSST site. \textit{Right:} bias and dispersion of colours due to airmass (point sizes) and PWV absorption (point colours) compared with out of atmosphere magnitude.}\label{fig:atm_params}
\end{figure}

The impact of atmosphere on magnitude colours can be estimated by comparing magnitudes measured below atmosphere with magnitudes measured by the same instrument but out of atmosphere ($T_{\mathrm{atm}}(\lambda)=1$).
Seasonal effects can affect $i-z$ colour up to 40~mmag, if the observatory site alternates with dry and wet seasons (Figure~\ref{fig:atm_params}). The PWV amplitude explored in Figure~\ref{fig:atm_params} is typical of the Rubin Observatory site according to the MERRA-2 database~\cite{MERRA2}. 

\begin{figure}[h!]
\centering
\includegraphics[width=0.315\textwidth]{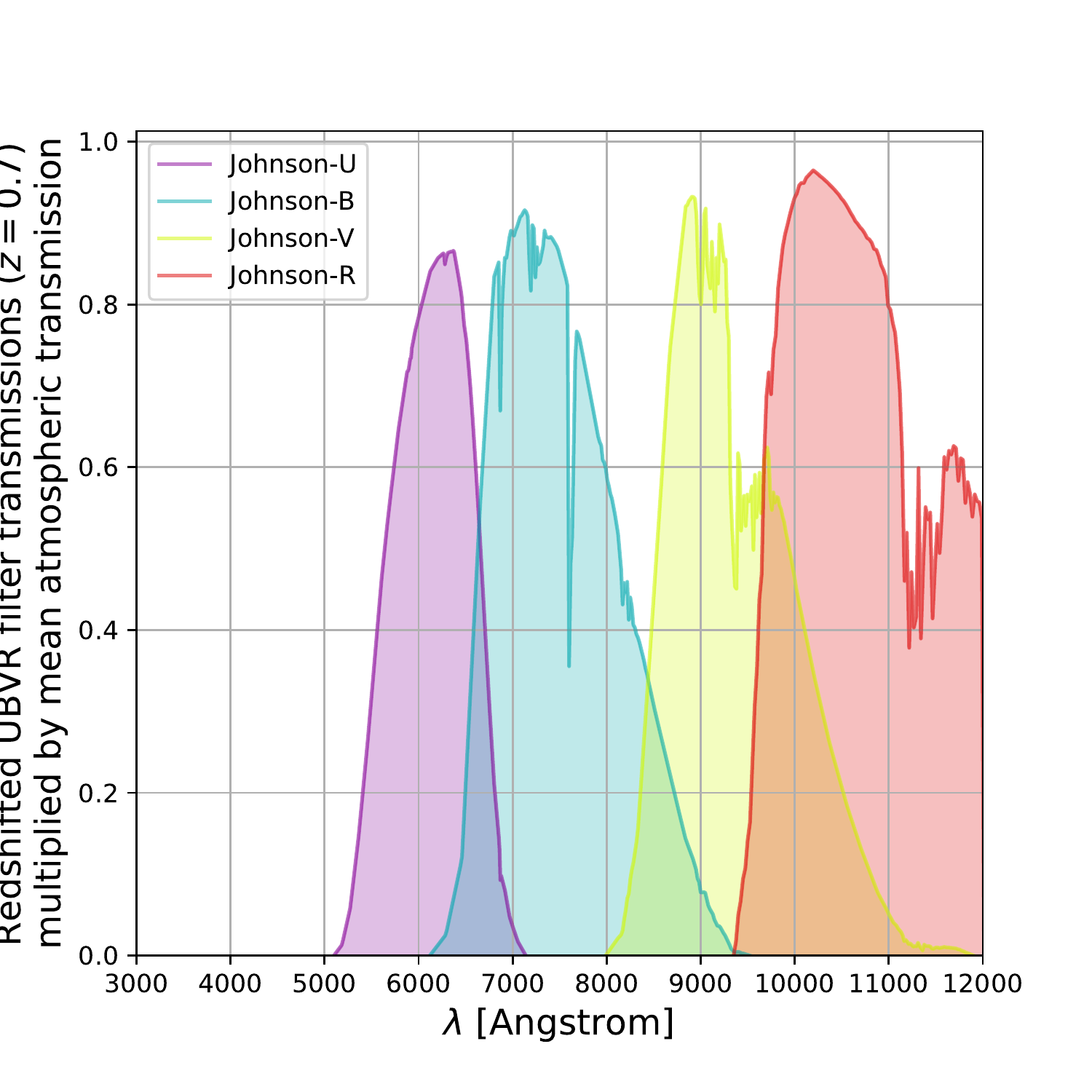}\hfill
\includegraphics[width=0.63\textwidth]{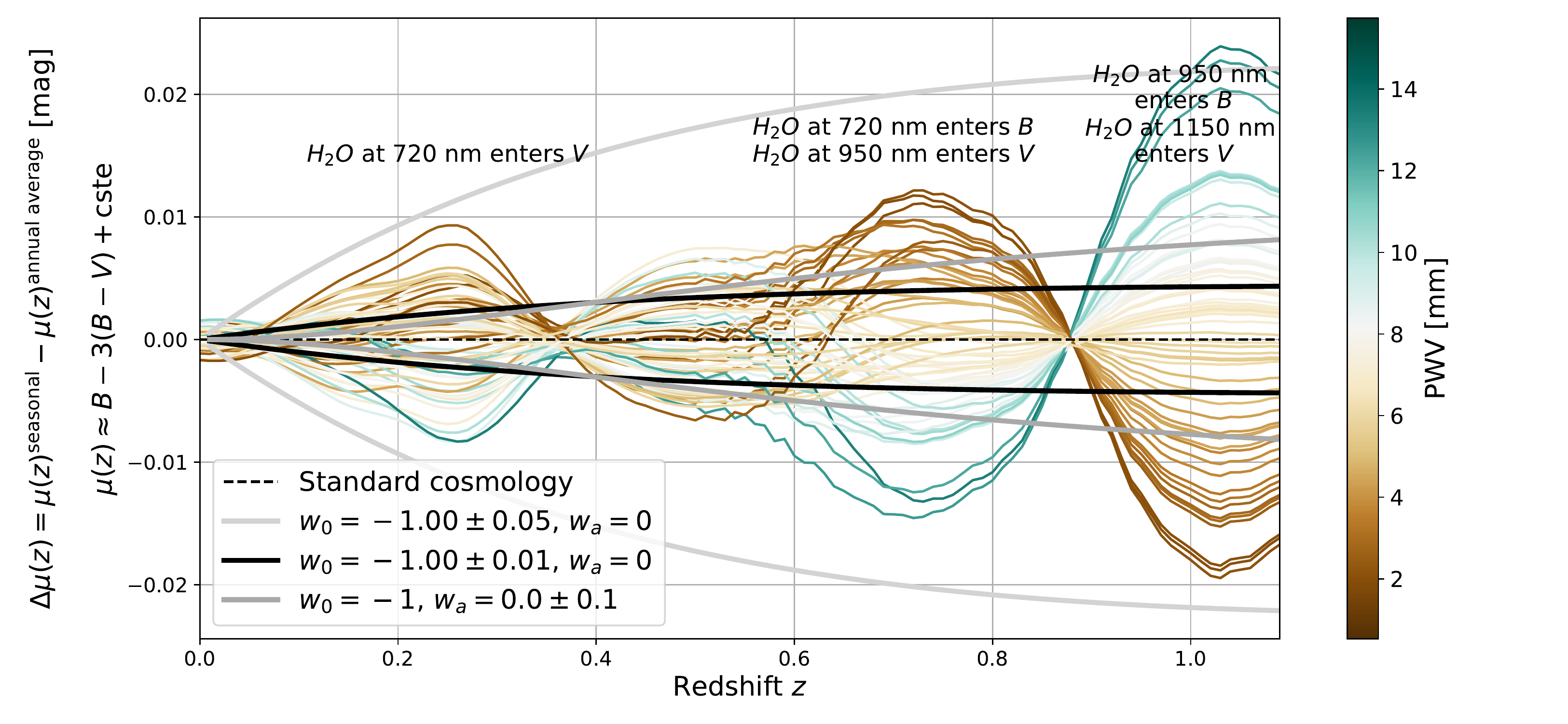}
\caption{\textit{Left:} effective redshifted UBVR filter throughput with a mean atmospheric transmission ($z=0.7$). \textit{Right:} impact of PWV on SNIa distance moduli with respect to a mean distance moduli obtained with an effective annual average atmosphere, and comparison to cosmological signatures.}\label{fig:atm_w}
\end{figure}
We simulated effective redshifted UBVR filter throughputs using different atmospheric conditions (Figure~\ref{fig:atm_w} left). 
Then we computed SNIa distance moduli $\mu(z) =B-3(B-V)$ shifts around airmass 1.2 for different PWV values with the effective redshifted filters. Variations around the mean Hubble diagram up to 10~mmag are seen around redshift $z\approx 0.7$ depending on season humidity, whereas a 1\% shift of $w_0$  from $-1$ leads to a 3~mmag offset at same redshift (Figure~\ref{fig:atm_w} right). This comparison helps setting requirements on the photometric calibration: to reach per-mille accuracy on the measurement of the dark energy equation of state parameter $w_0$ with SNe~Ia, on-site PWV must be measured seasonally at a millimetre precision approximately. 

The on-site PWV estimate can be performed using an auxiliary telescope equipped for spectrophotometry, like at the Rubin Observatory.Using a full forward modelling approach, the \texttt{Spectractor} software~\cite{spectractorascl} can extract the atmospheric transmission from slitless spectrophotometer exposures with accuracy, according to tests on simulations and real data. 

\end{document}